\documentclass{aa}

\usepackage{graphicx}
\usepackage{txfonts}
\usepackage{amsmath}
\usepackage{flafter}
\usepackage{lineno}
\usepackage{placeins}
\usepackage{flushend}
\usepackage[
  colorlinks=true,
  linkcolor=blue,
  citecolor=blue,
  urlcolor=blue,
  filecolor=blue
]{hyperref}

\newcommand{\neiiioii}{\ensuremath{[\mathrm{Ne\,III}]/[\mathrm{O\,II}]}}
\newcommand{\logneiiioii}{\ensuremath{\log([\mathrm{Ne\,III}]/[\mathrm{O\,II}])}}
\begin{document}

\title{Six-Class BPT Galaxy Classification for Survey-Scale AGN Candidate Prioritization: Deep Tabular Model and Informative Missingness Signals}

\titlerunning{Six-class BPT galaxy classification}
\authorrunning{S. Gui}

\author{Shanquan Gui\inst{1}}

\institute{Max Planck Institute for Astrophysics,
Karl-Schwarzschild-Str. 1, D-85741 Garching, Germany\\
\email{guishq25@mpa-garching.mpg.de}}

\date{Received xxx; accepted xxx}

\abstract
{The Baldwin--Phillips--Terlevich (BPT) diagram is widely used to classify galaxies into star-forming systems, composite galaxies, and active galactic nuclei (AGNs), but its survey-scale application is limited by the requirement for high signal-to-noise emission-line measurements.}
{We test whether machine-learning models can reproduce six-class BPT labels while using measured quantities, derived line ratios, and potentially informative missing-data patterns as inputs.}
{We analyze 1.47 million galaxies with a 27-dimensional feature set that combines raw survey measurements, derived quantities, and missingness indicators. Five deep tabular architectures are benchmarked against gradient-boosted trees and classical machine-learning baselines, and the resulting probabilities are evaluated through hard-classification metrics, precision--recall curves, top-$k$ retrieval, ablation tests, and feature-interpretation diagnostics.}
{CNN--Transformer gives the strongest overall classification performance (accuracy = 0.8266), while boosted trees remain highly competitive for this low-dimensional tabular problem. In the binary star-forming-versus-AGN comparison, CNN--Transformer achieves a Class~1 versus Class~4 ROC AUC of 0.9998. Missingness indicators provide substantial predictive information, especially the \texttt{OH\_P50\_missing} feature. Feature interpretation further shows that \logneiiioii, combined with stellar mass or specific star-formation rate, separates star-forming galaxies from AGN hosts.}
{The models are most useful as AGN candidate-ranking tools that complement, rather than replace, traditional BPT diagnostics. High-ranked samples can reach high purity, while broader candidate lists recover most AGNs, but transferability to other surveys requires further validation.}


\maketitle

%% From the front matter, we move on to the body of the paper.
%% Sections are demarcated by \section and \subsection, respectively.
%% Observe the use of the LaTeX \label
%% command after the \subsection to give a symbolic KEY to the
%% subsection for cross-referencing in a \ref command.
%% You can use LaTeX's \ref and \label commands to keep track of
%% cross-references to sections, equations, tables, and figures.
%% That way, if you change the order of any elements, LaTeX will
%% automatically renumber them.

\section{Introduction} 

The classification of galaxies based on optical emission-line ratios is a cornerstone of extragalactic astrophysics. The Baldwin--Phillips--Terlevich (BPT) diagnostic \citep{Baldwin1981,VeilleuxOsterbrock1987} uses the ratios of strong nebular emission lines to distinguish star-forming (SF) galaxies, composite systems, and active galactic nuclei (AGNs), including the Seyfert and LINER subclasses. These classifications provide a fundamental framework for studying galaxy evolution, supermassive black hole growth, and the co-evolution of star formation and nuclear activity. In its standard form, BPT classification relies on high signal-to-noise ratio (S/N) measurements of four diagnostic emission lines---[O~III]~$\lambda5007$, H$\beta$, [N~II]~$\lambda6584$, and H$\alpha$---and on empirically or theoretically motivated demarcation curves \citep{Kewley2001,Kauffmann2003,Kewley2006}.

Large spectroscopic surveys, most notably the Sloan Digital Sky Survey (SDSS), have made BPT-based classification central to statistical studies of nearby galaxies \citep{York2000,Abdurrouf2022,Almeida2023}. Despite this success, the diagnostic has an important practical limitation when applied to survey catalogs. A significant fraction of observed galaxies have one or more diagnostic emission lines below reliable detection thresholds, preventing robust placement on the BPT diagram. This incompleteness motivates complementary approaches capable of exploiting information beyond the four standard BPT line ratios, including host-galaxy properties and auxiliary spectral measurements. Related work has shown that additional spectral information can help distinguish AGN and star-forming systems when H$\alpha$ or [N~II] measurements are unavailable \citep{LevesqueRichardson2014,MunozSantos2025,TeimooriniaKeown2018}.

Machine-learning methods provide one such complementary route. Data-mining
and machine-learning approaches are now common in astronomy, particularly
for classification and ranking tasks in large surveys
\citep{BallBrunner2010,FlukeJacobs2020}. A recent study by \citet{Mazoochi2026} demonstrated the feasibility of
this strategy using classical algorithms, including decision trees,
random forests, support vector machines, and k-nearest neighbors, for
binary AGN versus star-forming galaxy classification. That result is an
important starting point, but it leaves three issues that are central for
survey-scale use. First, practical catalogs require the full multi-class
BPT taxonomy rather than only a binary AGN/star-forming split. Second,
modern deep tabular architectures should be compared against strong
boosted-tree baselines, since tabular benchmarks show that deep models do
not always dominate structured low-dimensional data
\citep{Gorishniy2021,Grinsztajn2022}. Third, missing measurements in
astronomical catalogs are not necessarily missing completely at random and
may encode observational selection or astrophysical information
\citep{Rubin1976,LittleRubin2002}.

In this paper, we perform a survey-scale six-class BPT classification
study using 1.47 million galaxies. We construct a 27-dimensional feature
set from original survey measurements, derived quantities, and binary
missingness indicators, and we benchmark five deep tabular models against
LightGBM, XGBoost, and classical machine-learning baselines. We then
evaluate class-wise performance, feature ablations, AGN
precision--recall and top-$k$ retrieval, and feature-importance trends.
CNN--Transformer gives the strongest overall classification performance
(accuracy = 0.8266), while boosted trees remain highly competitive for
this low-dimensional tabular problem. The binary Class~1 versus Class~4
comparison is much more accurate, with a ROC AUC of 0.9998. Missingness
indicators, especially \texttt{OH\_P50\_missing}, carry substantial
predictive information, and the predicted AGN probabilities are most
useful as candidate-ranking scores rather than contamination-free
replacements for traditional BPT diagnostics. Feature interpretation
further shows that
the \neiiioii\ line ratio, together with stellar mass or specific
star-formation rate, provides a physically meaningful separation between
star-forming galaxies and AGN hosts.

The remainder of this paper is organized as follows.
Section~\ref{sec:data_feature_pipeline} describes the dataset and
feature-construction pipeline. Section~\ref{sec:methods} presents the
model architectures, baselines, and training protocol.
Section~\ref{sec:results_discussion} reports the classification,
ablation, AGN-retrieval, and feature-interpretation results.
Section~\ref{sec:conclusions} summarizes the main conclusions.

%% The "ht!" tells LaTeX to put the figure "here" first, at the "top" next
%% and to override the normal way of calculating a float position.
%% The asterisk after "figure" tells the compiler to span multiple columns
%% if a two column style is selected.

\section{Data and Feature Pipeline}
\label{sec:data_feature_pipeline}

We use the MPA-JHU value-added galaxy catalog, which combines SDSS
spectroscopic redshifts and emission-line measurements with derived
physical quantities, including stellar masses, star-formation rates,
gas-phase metallicities, and BPT classifications. The catalog is based on
the SDSS Data Release~7 spectroscopic galaxy sample \citep{Abazajian2009};
the version used here was accessed through the SDSS MPA-JHU
galaxy-measurements interface\footnote{SDSS MPA-JHU galaxy
measurements: \url{https://www.sdss4.org/dr17/spectro/galaxy_mpajhu/}.}.
The raw table contains 1,843,200 galaxy records, and we use nine columns,
including redshift ($Z$), stellar mass (\texttt{LGM\_TOT\_P50}),
star-formation rate (\texttt{SFR\_TOT\_P50}), gas-phase metallicity
(\texttt{OH\_P50}), [Ne~III]$\lambda3869$ flux and uncertainty, [O~II]
flux and uncertainty, and the target BPT label (\texttt{BPTCLASS}).

\begin{figure}[!htbp]
\centering
\includegraphics[width=0.95\columnwidth,height=0.45\textheight,keepaspectratio]{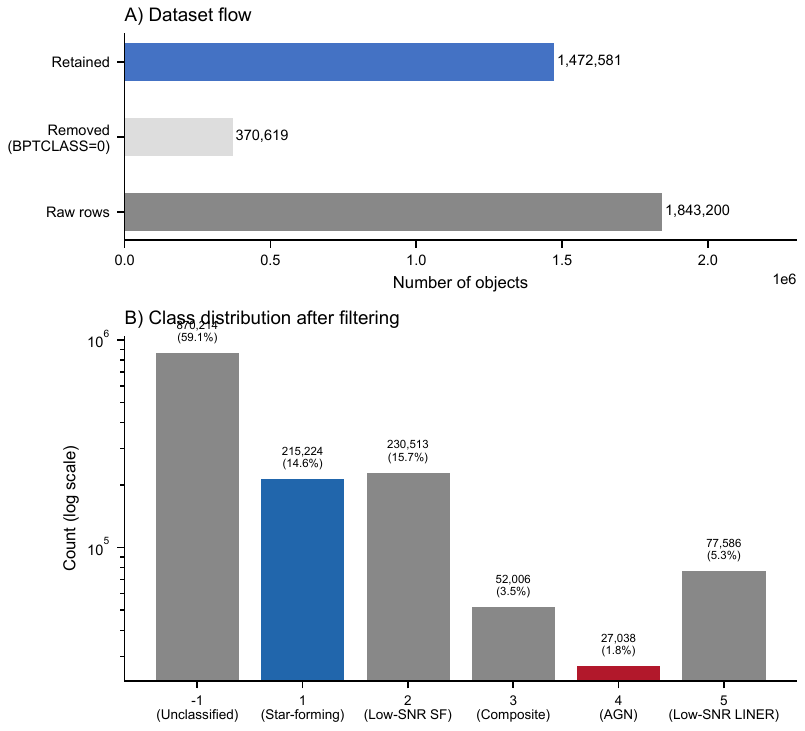}
\caption{
Distribution of the 1.47 million galaxies among the six BPT classes
used in this study: unclassified galaxies, star-forming galaxies,
low-S/N star-forming galaxies, composite galaxies, AGNs, and low-S/N
LINERs.
}
\label{fig:dataset_distribution}
\end{figure}

\begin{figure}[!htbp]
\centering
\includegraphics[width=0.95\columnwidth,height=0.45\textheight,keepaspectratio]{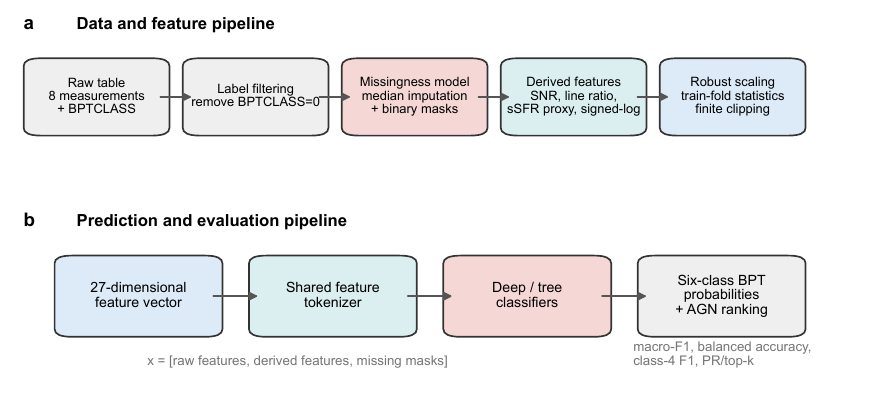}
\caption{
Overview of the data-processing, feature-construction, prediction,
and evaluation pipeline adopted in this work.
The upper panel illustrates the data and feature pipeline,
including label filtering, missing-value modeling, derived-feature
construction, and robust normalization.
The lower panel summarizes the prediction and evaluation workflow,
where the 27-dimensional feature vector is processed by deep-learning
or tree-based classifiers to produce six-class BPT probabilities and
AGN candidate rankings.
}
\label{fig:pipeline}
\end{figure}

After removing objects with \texttt{BPTCLASS=0},
the final dataset contains 1,472,581 galaxies
distributed across six valid classes. The raw catalog and the final
class subset are summarized in panels A and B of
Figure~\ref{fig:dataset_distribution}.
The complete data-processing, feature-construction, prediction, and
evaluation workflow is summarized in Figure~\ref{fig:pipeline}.

\subsection{Missing-Value Modeling}

Astronomical survey catalogs contain a substantial number of missing
measurements encoded as \texttt{-9999}.
Removing incomplete samples would alter the class distribution,
whereas replacing missing values with zeros would incorrectly treat
the absence of a measurement as a physical quantity.

We therefore adopt a median-imputation strategy combined with
binary missingness indicators.
For the $j$-th feature, the missingness mask is defined as
\begin{equation}
m_{ij}
=
\begin{cases}
1, & \text{if } x_{ij}\ \text{is missing},\\
0, & \text{otherwise}.
\end{cases}
\label{eq:missing_mask}
\end{equation}
Missing values are replaced using the median computed exclusively
from the training fold,
\begin{equation}
\tilde{x}_{ij}
=
\begin{cases}
x_{ij}, & \text{if observed},\\
\mathrm{median}_{j}, & \text{if missing}.
\end{cases}
\label{eq:median_imputation}
\end{equation}
The validation fold uses only preprocessing statistics derived
from the corresponding training fold in order to avoid information leakage.

The final model input contains both the imputed continuous features
and the binary missingness indicators,
\begin{equation}
\mathbf{x}
=
[
\tilde{\mathbf{x}},
\mathbf{m}
].
\end{equation}
This design enables the model to exploit not only the measured values
themselves but also patterns of data availability.

\subsection{Derived Features}

Eleven derived features are constructed from the eight original survey measurements. Signal-to-noise ratios are defined as
\begin{align}
{\rm SNR}_{\rm NeIII}
&=
\frac{F_{\rm NeIII}}
{\sigma_{\rm NeIII}+\epsilon},
\\
{\rm SNR}_{\rm OII}
&=
\frac{F_{\rm OII}}
{\sigma_{\rm OII}+\epsilon},
\end{align}
where $F_{\rm NeIII}$ denotes the flux of
[\ion{Ne}{3}] $\lambda3869$,
and
\begin{equation}
F_{\rm OII}
=
F_{3726}
+
F_{3729}
\end{equation}
is the total flux of the
[\ion{O}{2}] $\lambda\lambda3726,3729$
doublet.
Here $\sigma_{\rm NeIII}$ and $\sigma_{\rm OII}$ denote the
corresponding flux uncertainties, which have been scaled according to
the MPA-JHU DR7 raw-data documentation\footnote{\url{https://wwwmpa.mpa-garching.mpg.de/SDSS/DR7/raw_data.html}},
and $\epsilon$ is a small constant introduced for numerical stability. An ionization-sensitive line-ratio feature is computed as
\begin{equation}
R_{\rm NeIII/OII}
=
\frac{F_{\rm NeIII}}
{F_{\rm OII}+\epsilon}.
\end{equation}
A proxy for the specific star-formation rate is defined as
\begin{equation}
\log {\rm sSFR}_{\rm proxy}
=
{\rm SFR\_TOT\_P50}
-
{\rm LGM\_TOT\_P50},
\label{eq:ssfr_proxy}
\end{equation}
where \texttt{SFR\_TOT\_P50} and \texttt{LGM\_TOT\_P50}
represent the logarithmic star-formation rate and stellar mass,
respectively.

To reduce the dynamic range of fluxes, uncertainties,
signal-to-noise ratios, and line-ratio quantities,
we apply a signed logarithmic transformation,
\begin{equation}
\phi(x)
=
{\rm sign}(x)
\log(1+|x|).
\end{equation}
The final feature representation consists of
eight original measurements,
eleven derived quantities,
and eight missingness indicators,
yielding a total of 27 input dimensions.

\section{Methods}
\label{sec:methods}

\subsection{Problem Formulation}

We formulate survey-scale BPT identification as a six-class tabular classification problem.
Given a dataset
\begin{equation}
\mathcal{D}
=
\left\{
(\mathbf{x}_i, y_i)
\right\}_{i=1}^{N},
\end{equation}
where $\mathbf{x}_i \in \mathbb{R}^{d}$ denotes the feature vector of the $i$th galaxy and
$y_i$ represents its BPT label, the objective is to learn a parametric mapping
\begin{equation}
f_{\theta} :
\mathbb{R}^{d}
\rightarrow
[0,1]^{C},
\end{equation}
where $C=6$ is the number of target classes. The valid label set is
\begin{equation}
\mathcal{Y}
=
\{-1,1,2,3,4,5\},
\end{equation}

corresponding to the six valid catalog labels used in this work.
Objects with \texttt{BPTCLASS=0} are excluded from training because their
physical interpretation is ambiguous in the current catalog.

For each galaxy, the model outputs a probability vector
\begin{equation}
\hat{\mathbf{p}}_i
=
f_{\theta}(\mathbf{x}_i),
\end{equation}
with predicted class
\begin{equation}
\hat{y}_i
=
\arg\max_{c}
\hat{p}_{i,c}.
\end{equation}
Although the optimization objective is six-class classification,
the primary astrophysical focus of this work is the distinction
between star-forming galaxies (Class~1) and AGNs (Class~4).
Therefore, in addition to standard multiclass metrics,
we report class-specific precision, recall, and F1 scores,
together with one-vs-rest ROC AUC,
binary Class~1 versus Class~4 AUC,
and candidate-ranking metrics relevant to AGN retrieval.

\subsection{Feature Tokenization}

With the exception of MLP--ResNet, all deep architectures share a common
feature-tokenization module, following the feature-token strategy used in
recent deep tabular models \citep{Huang2020,Gorishniy2021}. Given an input vector
\begin{equation}
\mathbf{x}
=
[x_1,x_2,\ldots,x_d],
\end{equation}
each scalar feature is projected into a learnable embedding space through
\begin{equation}
\mathbf{t}_j
=
x_j \mathbf{w}_j + \mathbf{b}_j ,
\end{equation}
where
$\mathbf{w}_j,\mathbf{b}_j \in \mathbb{R}^{D}$
are learnable parameters and $D$ denotes the token dimension. The resulting token sequence is
\begin{equation}
\mathbf{T}
=
[\mathbf{t}_1,\mathbf{t}_2,\ldots,\mathbf{t}_d]
\in
\mathbb{R}^{d\times D}.
\end{equation}

This representation serves two purposes.
First, it projects heterogeneous physical measurements into a common
learnable feature space.
Second, it preserves feature-level structure, enabling convolution,
attention, and graph-based modules to explicitly model interactions
between input variables.

\subsection{Deep Tabular Architectures}

\begin{figure}[!htbp]
\centering
\includegraphics[width=0.95\columnwidth,height=0.45\textheight,keepaspectratio]{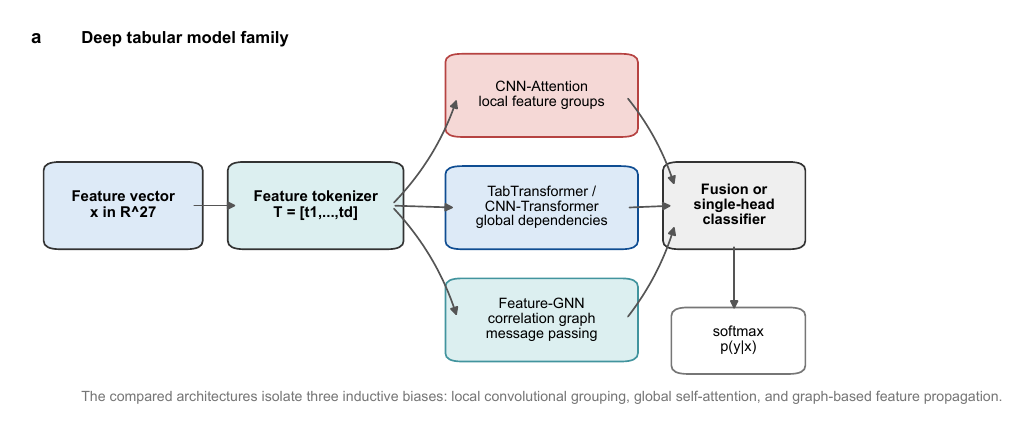}
\caption{
Overview of the deep tabular architectures evaluated in this work.
The models include MLP--ResNet, CNN--Attention, CNN--Transformer,
Feature-GNN, and Deep Hybrid.
Each architecture receives the same 27-dimensional feature representation
and learns predictive mappings for six-class BPT classification.
}
\label{fig:model_architecture}
\end{figure}

The relationships among the deep tabular architectures investigated in
this work are shown in Figure~\ref{fig:model_architecture}. The purpose
of comparing these models is not to increase architectural complexity,
but rather to isolate and evaluate three distinct forms of inductive
bias: local compositional structure captured by convolutional operations,
global feature dependencies modeled through self-attention mechanisms,
and graph-based message passing over feature-correlation networks.

\subsubsection{MLP--ResNet}

MLP--ResNet serves as the baseline deep-learning architecture, using
residual connections motivated by residual neural networks
\citep{He2016,Gorishniy2021}. Given an input vector $\mathbf{x}$, an initial hidden representation is
constructed using a linear projection followed by batch normalization and
GELU activation,
\begin{equation}
\mathbf{h}_0
=
{\rm GELU}
\left(
{\rm BN}
(W_0\mathbf{x}+b_0)
\right).
\end{equation}
The network then stacks four residual blocks, which can be written as
\begin{equation}
\begin{aligned}
\mathbf{u}_{l}
&=
{\rm GELU}
\left[
{\rm BN}(W_{l,1}\mathbf{h}_l+b_{l,1})
\right],\\
\mathbf{r}_{l}
&=
{\rm BN}
\left[
W_{l,2}{\rm Dropout}(\mathbf{u}_{l})
\right],\\
\mathbf{h}_{l+1}
&=
\mathbf{h}_{l}+\mathbf{r}_{l}.
\end{aligned}
\end{equation}
The final hidden representation is mapped to six-class logits through
a linear classification head.

\subsubsection{CNN--Attention}

The CNN--Attention architecture first processes the feature-token sequence
using three one-dimensional convolutional layers in order to capture
local feature combinations, followed by attention-based pooling. Given convolutional outputs $\mathbf{h}_j$,
attention scores are computed as
\begin{equation}
s_j
=
\mathbf{w}_a^{\top}\mathbf{h}_j ,
\end{equation}
with normalized attention weights
\begin{equation}
a_j
=
\frac{\exp(s_j)}
{\sum_k \exp(s_k)}.
\end{equation}
The sample-level representation is obtained through weighted aggregation,
\begin{equation}
\mathbf{z}_{\rm cnn}
=
\sum_{j=1}^{d}
a_j \mathbf{h}_j.
\end{equation}
This architecture assumes that fluxes, uncertainties, signal-to-noise
ratios, line ratios, and missingness indicators may contribute through
localized feature combinations.

\subsubsection{TabTransformer and CNN--Transformer}

TabTransformer models global dependencies among all feature tokens through
multi-head self-attention \citep{Vaswani2017,Huang2020}. Given query, key, and value matrices
$\mathbf{Q}$,
$\mathbf{K}$,
and
$\mathbf{V}$,
self-attention is defined as
\begin{equation}
{\rm Attention}
(\mathbf{Q},\mathbf{K},\mathbf{V})
=
{\rm softmax}
\left(
\frac{\mathbf{Q}\mathbf{K}^{\top}}
{\sqrt{D}}
\right)
\mathbf{V}.
\end{equation}
The final CLS-token representation is used as the sample embedding. The CNN--Transformer architecture augments this framework by introducing
local convolutional processing before the Transformer encoder.
Two one-dimensional convolutional layers generate local interaction
features,
\begin{equation}
\mathbf{U}
=
{\rm CNN}
(\mathbf{T}),
\end{equation}
which are subsequently processed by a three-layer Transformer encoder,
\begin{equation}
\mathbf{H}
=
{\rm TransformerEncoder}
(\mathbf{U}).
\end{equation}
A gated-attention pooling layer produces the final sample representation,
\begin{align}
\beta_j
&=
\frac{
\exp(\mathbf{w}_p^{\top}\mathbf{h}_j)
}
{
\sum_k
\exp(\mathbf{w}_p^{\top}\mathbf{h}_k)
},
\\
\mathbf{z}_{\rm att}
&=
\sum_j
\beta_j \mathbf{h}_j.
\end{align}
This design combines local feature composition with global attention-based
dependency modeling.

\subsubsection{Feature-GNN}

Feature-GNN treats each input feature as a graph node, adopting the general
message-passing perspective of graph neural networks \citep{KipfWelling2017}. A feature-correlation graph is constructed using the absolute Pearson
correlation matrix estimated from the training set,
\begin{equation}
A_{ij}
=
\left|
{\rm corr}(X_i,X_j)
\right|.
\end{equation}
Edges below a predefined threshold are removed, diagonal elements are set
to unity, and the adjacency matrix is row-normalized,
\begin{equation}
\tilde{A}_{ij}
=
\frac{A_{ij}}
{\sum_k A_{ik} + \epsilon}.
\end{equation}
Given token representations $\mathbf{T}$,
two rounds of graph message passing are applied,
\begin{align}
\mathbf{H}^{(1)}
&=
{\rm GELU}\!\left(
\tilde{\mathbf{A}}
\mathbf{T}
W_1
+
b_1
\right),
\\
\mathbf{H}^{(2)}
&=
{\rm GELU}\!\left(
\tilde{\mathbf{A}}
\mathbf{H}^{(1)}
W_2
+
b_2
\right).
\end{align}
The resulting node representations are aggregated using gated pooling
to obtain a graph-level embedding.

\subsubsection{Deep Hybrid}

Deep Hybrid architecture integrates the CNN--Attention,
TabTransformer, and Feature-GNN models in parallel.
Given an input feature vector $\mathbf{x}$,
the three branches independently generate latent representations
\begin{align}
\mathbf{z}_{\rm cnn}
&=
{\rm CNNAttention}(\mathbf{x}),
\\
\mathbf{z}_{\rm tr}
&=
{\rm TabTransformer}(\mathbf{x}),
\\
\mathbf{z}_{\rm gnn}
&=
{\rm FeatureGNN}(\mathbf{x}).
\end{align}
The outputs are concatenated to form a fused representation,
\begin{equation}
\mathbf{z}_{\rm hyb}
=
[
\mathbf{z}_{\rm cnn};
\mathbf{z}_{\rm tr};
\mathbf{z}_{\rm gnn}
].
\end{equation}
The fused feature vector is subsequently processed by a multilayer
perceptron classifier to produce the final six-class probability
distribution. 

By combining convolution-based local feature extraction,
attention-based global dependency modeling,
and graph-based feature interaction learning,
the Deep Hybrid model aims to exploit complementary inductive biases
for astronomical tabular data. A two-layer multilayer perceptron is then used to generate the final classification logits. The motivation behind this architecture is that BPT classification may
depend simultaneously on local spectral-feature combinations, global
cross-feature dependencies, and correlation-driven feature propagation.

\subsection{Classical Baselines}

To facilitate comparison with previous studies, we reproduce several
classical machine-learning models, including decision trees,
random forests \citep{Breiman2001}, k-nearest neighbors (KNN)
\citep{CoverHart1967}, and support vector classifiers with radial basis
function kernels (SVC-RBF) \citep{CortesVapnik1995}.

We additionally include logistic regression, XGBoost, and LightGBM,
which serve as strong tabular-learning baselines
\citep{ChenGuestrin2016,Ke2017,Grinsztajn2022}. All baseline models use the same 27-dimensional feature representation
and the same five-fold cross-validation protocol.

\subsection{Optimization and Training Protocol}

The primary experiments employ weighted cross-entropy loss. Given model logits $\mathbf{o}_i$,
class probabilities are computed through
\begin{equation}
\hat{p}_{i,c}
=
\frac{\exp(o_{i,c})}
{\sum_{k=1}^{C}
\exp(o_{i,k})}.
\end{equation}
The weighted cross-entropy loss is
\begin{equation}
\mathcal{L}_{\rm CE}
=
-\frac{1}{N}
\sum_{i=1}^{N}
w_{y_i}
\log
\hat{p}_{i,y_i},
\end{equation}
where $w_c$ denotes the class weight estimated from the training-fold
class distribution. For AGN-focused experiments, we additionally evaluate focal loss
\citep{Lin2017},
\begin{equation}
\mathcal{L}_{\rm focal}
=
-\frac{1}{N}
\sum_{i=1}^{N}
(1-\hat{p}_{i,y_i})^{\gamma}
w_{y_i}
\log
\hat{p}_{i,y_i},
\end{equation}

with $\gamma=2$.

All experiments use stratified five-fold cross-validation.
Preprocessing statistics, including median-imputation values,
RobustScaler parameters, and class weights, are estimated exclusively
from the training folds to prevent information leakage. Deep models are optimized using AdamW \citep{KingmaBa2015,LoshchilovHutter2019}
with a learning rate of
$5\times10^{-4}$, weight decay of $10^{-4}$,
batch size of 8192, gradient clipping threshold of 5.0,
and a maximum of 100 training epochs.
Early stopping with a patience of six epochs is applied. Model selection is based on
\begin{equation}
{\rm Score}
=
{\rm MacroF1}
+
{\rm F1}_{\rm Class4},
\end{equation}
which simultaneously encourages strong overall multiclass performance
and robust AGN identification. To reduce computational cost and mitigate domination by majority classes,
a maximum of 50,000 training samples per class is retained within each
training fold.

\section{Results and Discussion}
\label{sec:results_discussion}

This section presents the main empirical results of the study. We first
compare the overall and class-wise performance of the evaluated models,
including convergence behavior and ablation tests. We then examine AGN
precision--recall and top-$k$ retrieval performance, followed by
feature-importance and diagnostic-plane analyses that connect the model
predictions to physically interpretable quantities.

All experiments were conducted using stratified five-fold
cross-validation, ensuring that the class distribution was preserved
within each fold. For each split, models were trained on four folds and
evaluated on the remaining fold.

\subsection{Model Performance, AGN Retrieval, and Ablation}

\begin{figure*}[!htbp]
\centering
\includegraphics[width=0.86\textwidth,height=0.55\textheight,keepaspectratio]{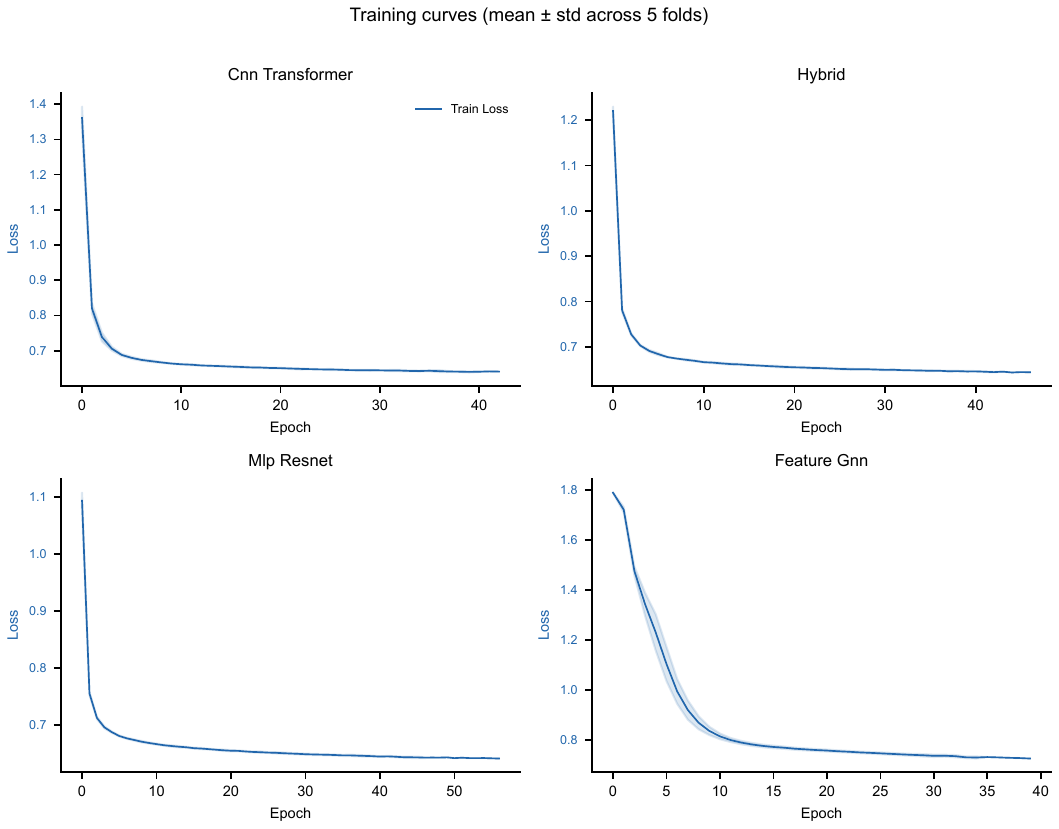}
\caption{
Training and validation learning curves for the deep tabular models
evaluated in this work. Curves show the evolution of the optimization
objective and classification performance as a function of training
epoch.
}
\label{fig:training_curves}
\end{figure*}

For the deep-learning models, training and validation learning curves are examined to assess convergence behavior and potential overfitting
(Figure~\ref{fig:training_curves}). Most architectures converge within
the first few tens of epochs and
exhibit stable validation performance thereafter, indicating that the
optimization procedure is well behaved. 

\begin{figure*}[!htbp]
\centering
\includegraphics[width=0.86\textwidth,height=0.55\textheight,keepaspectratio]{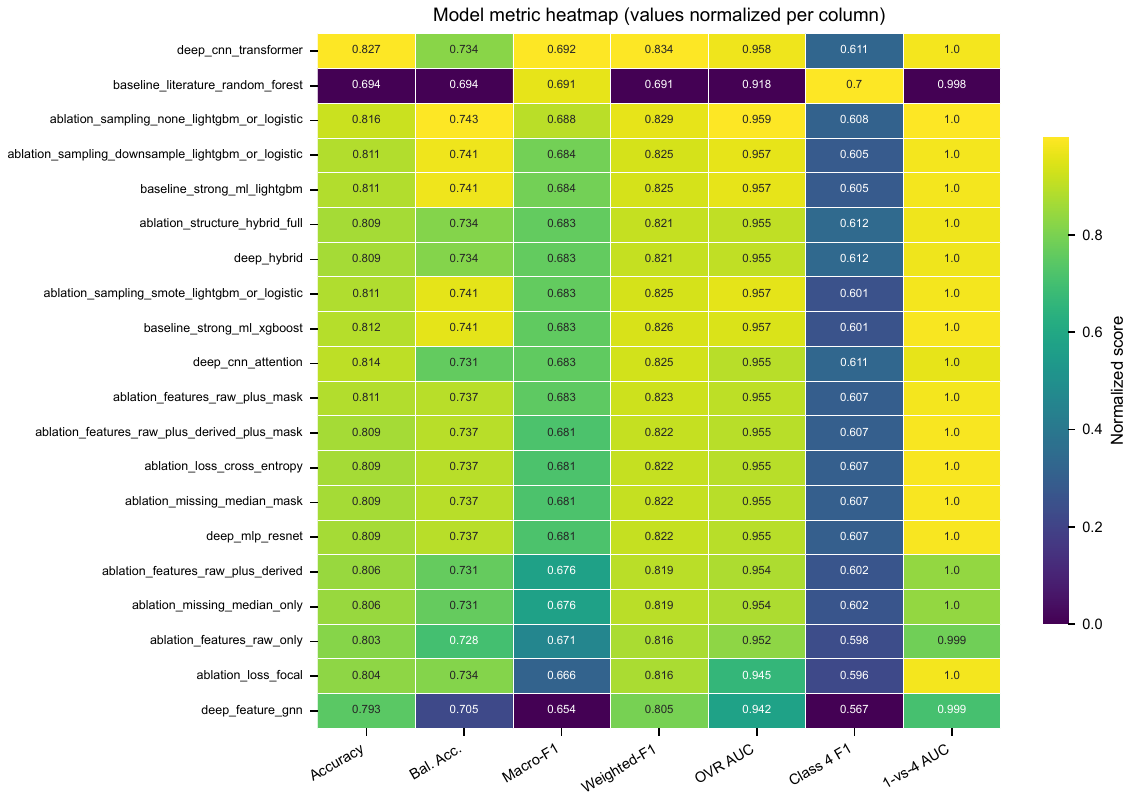}
\caption{
Normalized performance heatmap for all evaluated models across multiple
evaluation metrics, including accuracy, balanced accuracy, macro-F1,
weighted F1, one-vs-rest ROC AUC, and AGN-specific metrics. Darker
colors indicate stronger performance. 
}
\label{fig:metric_heatmap}
\end{figure*}

We first compare the full set of evaluated models using the principal
classification metrics. The normalized metric heatmap in Figure~\ref{fig:metric_heatmap} summarizes
the broad performance trends across the evaluated models.

Among all models, CNN--Transformer achieves the highest macro-F1 score,
reaching $0.6925 \pm 0.0048$. The same model attains an overall
accuracy of 0.8266, a weighted F1 score of 0.8342, and a multiclass
one-vs-rest AUC of 0.9579. However, the performance advantage over
strong tree-based methods is modest: LightGBM reaches a macro-F1 score
of $0.6842 \pm 0.0010$ and an OVR AUC of 0.9574, and XGBoost produces
similar results. 
The class-wise results shown in the last column in Figure~\ref{fig:metric_heatmap} clarify where the models succeed and where they
fail. Class~1 (star-forming galaxies) is effectively separated from the
other classes: CNN--Transformer reaches a precision of 0.991, recall of
0.985, and F1 score of 0.988, with LightGBM yielding nearly identical
performance. By contrast, Class~4 (AGNs) remains substantially more
challenging. CNN--Transformer obtains a precision of 0.545, recall of
0.695, and F1 score of 0.611, implying that the models recover many AGNs
but include contamination from neighboring classes. Class~5 (low-S/N
LINERs) is the most difficult category, with F1 scores below 0.50 for the
best models, while Classes~2 and 3 occupy an intermediate difficulty
regime.

\begin{figure}[!htbp]
\centering
\includegraphics[width=0.95\columnwidth,height=0.45\textheight,keepaspectratio]{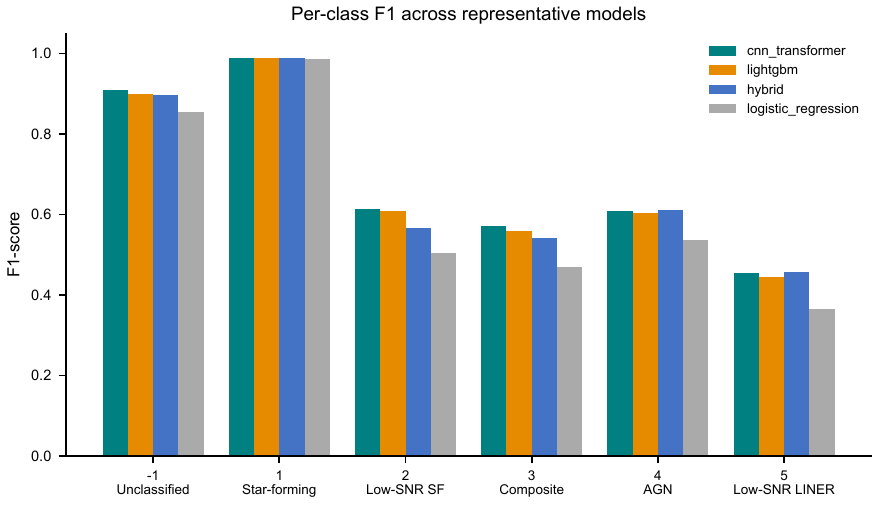}
\caption{
Per-class F1 scores for four evaluated models across the six BPT
categories. 
}
\label{fig:per_class_f1}
\end{figure}

Figure~\ref{fig:per_class_f1} compares the per-class F1 scores of four
representative architectures.
\begin{figure}[!htbp]
\centering
\includegraphics[width=0.95\columnwidth,height=0.45\textheight,keepaspectratio]{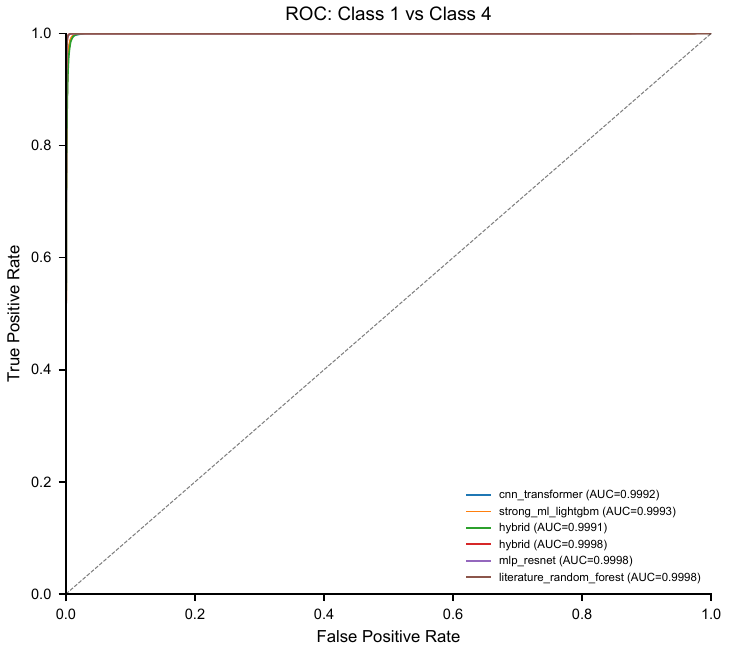}
\caption{
Receiver operating characteristic (ROC) curves for the binary
separation between Class~1 (star-forming galaxies) and Class~4
(AGNs). All evaluated models achieve near-perfect discrimination,
with AUC values approaching unity.
}
\label{fig:roc_class1_vs_class4}
\end{figure}
\begin{figure*}[!htbp]
\centering
\includegraphics[width=0.86\textwidth,height=0.55\textheight,keepaspectratio]{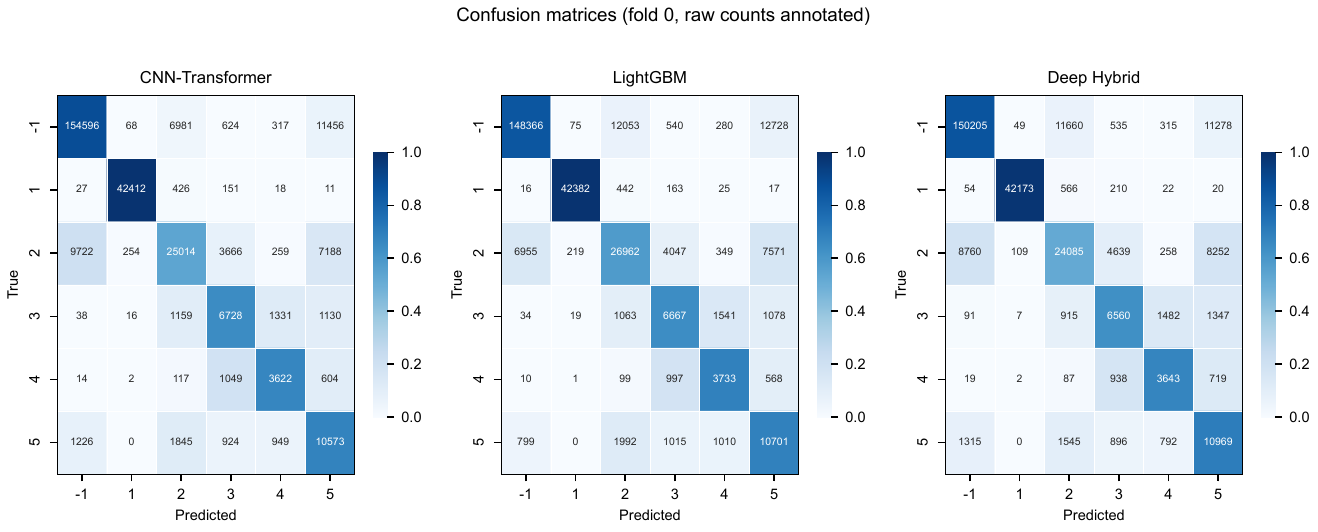}
\caption{
Row-normalized confusion matrices for representative models, including
CNN--Transformer, LightGBM, and Deep Hybrid. Each row is normalized by
the number of true objects in the corresponding class. AGN classification
errors are dominated by confusion with Class~3 (Composite) and Class~5
(low-S/N LINERs) rather than with Class~1 (star-forming galaxies).
}
\label{fig:confusion_matrices}
\end{figure*}
Star-forming galaxies (Class~1) achieve the highest F1 scores across
all models, indicating that this population is readily separable from
the remaining classes. In contrast, composite galaxies (Class~3),
AGNs (Class~4), and low-S/N LINERs (Class~5) exhibit lower F1 scores,
reflecting significant overlap near BPT classification boundaries.
Figure~\ref{fig:per_class_f1} further illustrates that differences among
the strongest models are generally modest, with CNN--Transformer,
LightGBM, and XGBoost achieving comparable performance across most
classes.

The receiver operating characteristic (ROC) curves for the binary
separation between Class~1 (star-forming galaxies) and Class~4
(AGNs) are shown in Figure~\ref{fig:roc_class1_vs_class4}. The very high binary AUC indicates that star-forming galaxies and AGNs
occupy largely distinct regions of the learned probability space, even
though the full six-class task remains more difficult. Thus, the main
classification errors arise near neighboring BPT boundaries---especially
between AGNs, composite galaxies, and low-S/N LINERs---rather than
between AGNs and star-forming galaxies.

Figure~\ref{fig:confusion_matrices} presents row-normalized confusion
matrices for three representative models.AGN misclassifications predominantly flow toward Class~3
(Composite) and Class~5 (low-S/N LINERs), rather than toward Class~1
(star-forming galaxies). In Fold~0, for example, only two AGNs are
misclassified directly as star-forming galaxies, whereas substantially
larger numbers are assigned to adjacent boundary classes. This behavior
indicates that the models recover the broad geometry of the BPT diagram:
the dominant failures occur near physically adjacent decision boundaries
rather than between fundamentally distinct ionization regimes.

\begin{figure*}[!htbp]
\centering
\includegraphics[width=0.86\textwidth,height=0.55\textheight,keepaspectratio]{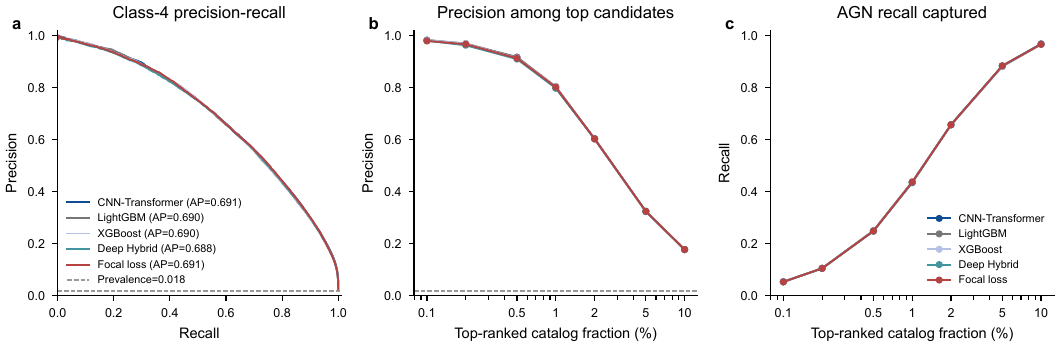}
\caption{
AGN retrieval performance evaluated using precision--recall curves and
top-$k$ candidate analysis. The left panel presents precision--recall
curves for representative models under the imbalanced survey
distribution. The middle panel shows precision among top-ranked
candidates as a function of the selected catalog fraction. The right
panel shows the corresponding AGN recall captured as the candidate list
is expanded according to predicted AGN probability.
}
\label{fig:agn_retrieval}
\end{figure*}

Because AGNs represent only 1.8\% of the full dataset, ROC AUC alone can
overestimate practical catalog-building performance. We therefore
evaluate AGN retrieval using precision--recall curves and top-$k$
candidate analyses (Figure~\ref{fig:agn_retrieval}). CNN--Transformer
achieves an average precision (AP) of 0.691, while LightGBM reaches
0.690. When galaxies are ranked by predicted AGN probability, the top
0.1\% of candidates reaches a precision of approximately 0.984. At the
top 1\% threshold, precision remains near 0.80 while recall reaches
approximately 0.44. Expanding the candidate pool to the top 10\% recovers
nearly 97\% of all AGNs in the sample.

These retrieval results show that the model outputs are most useful as a
ranking tool. Although a single hard threshold does not produce a
contamination-free AGN catalog, probability-based ranking can greatly
reduce the observational effort required for follow-up target selection.
This interpretation is more robust than treating the six-class predictions
as final physical classifications for every object.

\begin{figure}[!htbp]
\centering
\includegraphics[width=0.95\columnwidth,height=0.45\textheight,keepaspectratio]{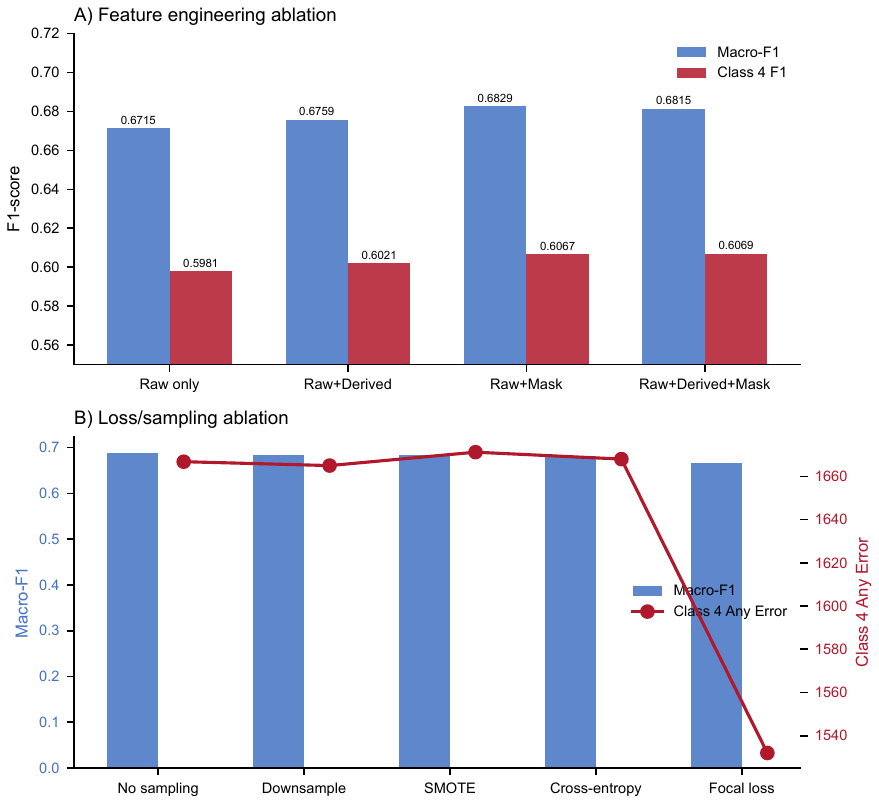}
\caption{
Summary of the ablation experiments. Panel (a) compares the contributions
of raw features, derived astrophysical features, and missingness
indicators. Panel (b) compares sampling strategies and loss functions. 
}
\label{fig:ablation_summary}
\end{figure}

We next examine which input groups are responsible for the performance.
Figure~\ref{fig:ablation_summary} summarizes the ablation experiments.
Using only the eight original survey measurements yields a macro-F1
score of 0.671. Adding derived astrophysical features increases the
score to 0.676, a gain of only 0.0044. In contrast, adding missingness
indicators without derived quantities raises macro-F1 to 0.683, an
improvement of 0.0114. The complete feature set achieves a macro-F1
score of 0.681. The same trend appears for the AGN class: the Class~4
F1 score increases from 0.598 for the raw-feature baseline to 0.602
after adding derived features and to 0.607 after incorporating
missingness indicators. Including both derived features and missingness
indicators does not further improve the result, suggesting partial
information overlap between the two feature groups.

Class-balancing strategies show a complementary trade-off. In the
sampling/loss experiments, the no-resampling LightGBM/logistic
configuration attains a balanced accuracy of 0.743 and a macro-F1 score
of 0.688. Random downsampling and SMOTE oversampling \citep{Chawla2002}
provide limited additional benefit. Focal loss behaves differently:
although macro-F1 decreases from 0.681 to 0.666, Class~4 recall rises
from 0.692 to 0.717. Thus, focal loss is useful when AGN completeness is
prioritized, whereas cross-entropy and gradient-boosted trees remain
preferable for balanced six-class classification.

\subsection{Feature Interpretation}

We then investigate which inputs drive the separation between star-forming
galaxies and AGN hosts. The purpose of this analysis is not to assign a
unique causal interpretation to each feature, because several quantities
are correlated and some are produced by the same survey pipeline. Instead,
we ask whether the dominant model signals are astrophysically meaningful,
whether catalog missingness contains reproducible information, and whether
the resulting separation is consistent with known AGN-host properties.

\begin{figure*}[!htbp]
\centering
\includegraphics[width=0.86\textwidth,height=0.55\textheight,keepaspectratio]{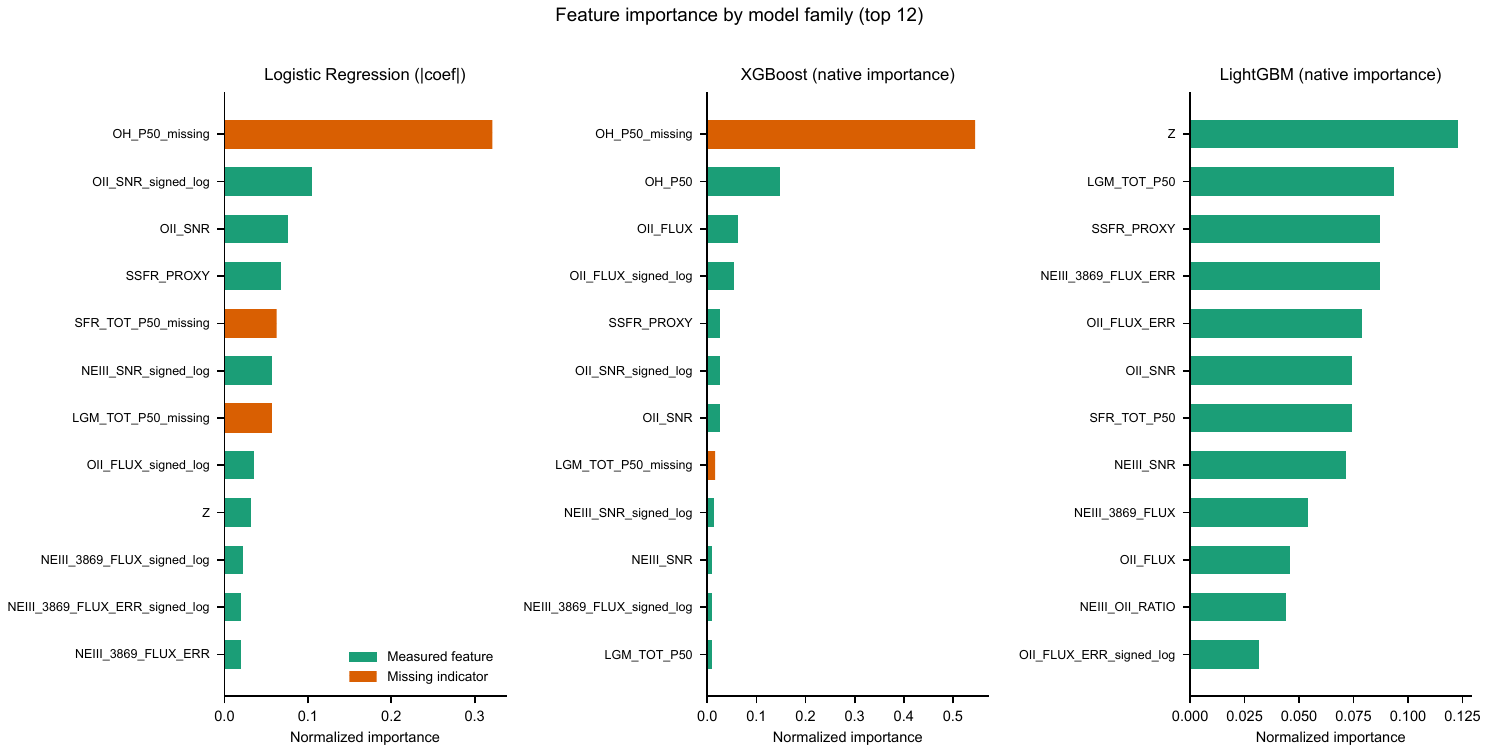}
\caption{
Feature-importance rankings derived from three complementary model
families: Logistic Regression, XGBoost, and LightGBM.
}
\label{fig:feature_importance}
\end{figure*}

Figure~\ref{fig:feature_importance} first summarizes the relative
importance of the input features across three complementary model
families.
The feature-importance rankings in Figure~\ref{fig:feature_importance}
show that the classifiers do not depend on a single diagnostic variable.
High-ranking inputs include host-galaxy properties, emission-line ratios,
and binary missingness indicators. The repeated appearance of
\texttt{OH\_P50\_missing} among the most informative features is
particularly important. It shows that the availability of catalog
measurements can itself help separate classes, consistent with the
ablation tests in Figure~\ref{fig:ablation_summary}, where missingness
indicators provide a larger performance gain than the derived features
alone.

\begin{figure*}[!htbp]
\centering
\includegraphics[width=0.86\textwidth,height=0.55\textheight,keepaspectratio]{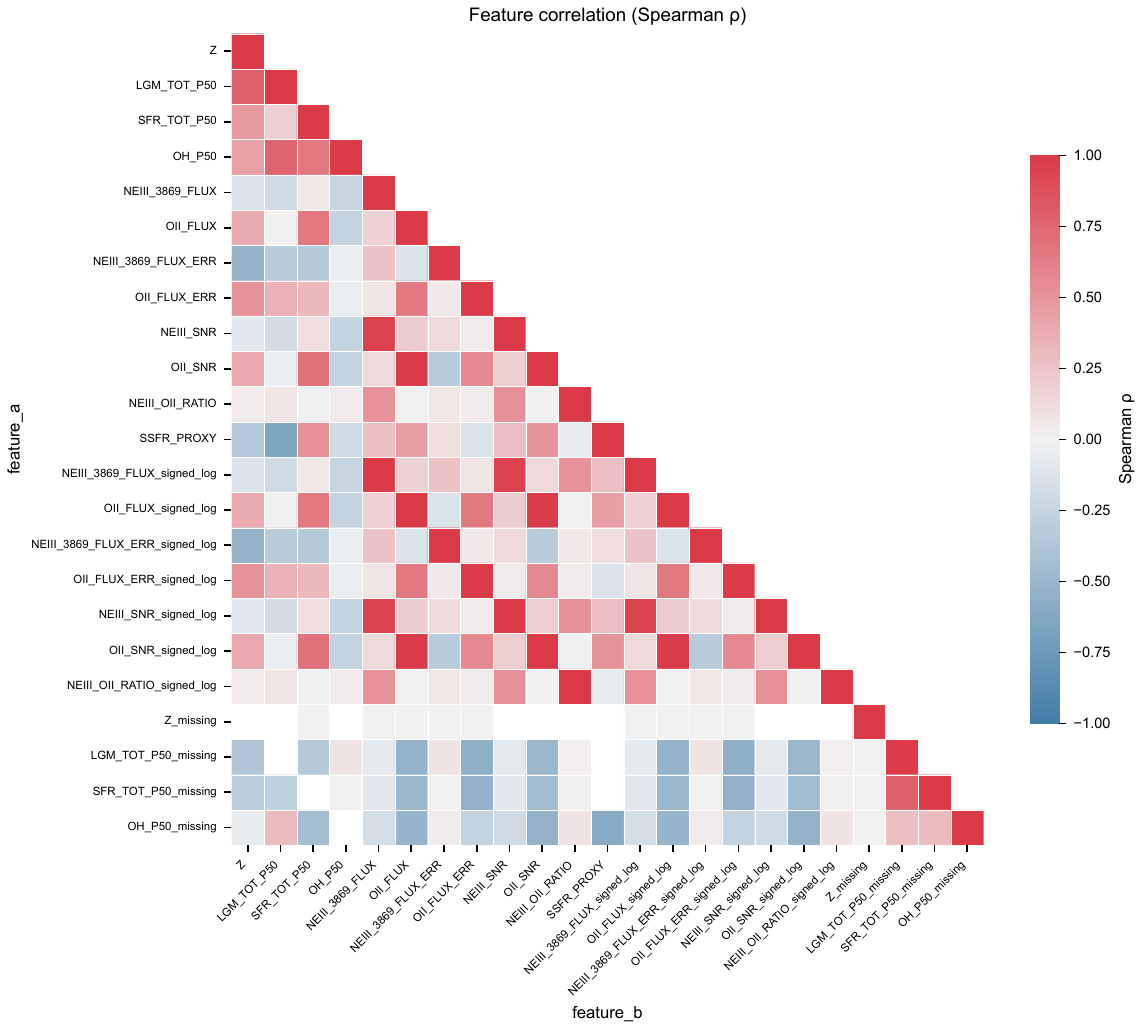}
\caption{
Spearman rank correlation matrix for all input features used in the
classification models.
}
\label{fig:feature_correlation}
\end{figure*}

Figure~\ref{fig:feature_correlation} shows the feature-correlation
structure that should be kept in mind when interpreting these rankings.
\begin{figure*}[!htbp]
\centering
\includegraphics[width=0.86\textwidth,height=0.55\textheight,keepaspectratio]{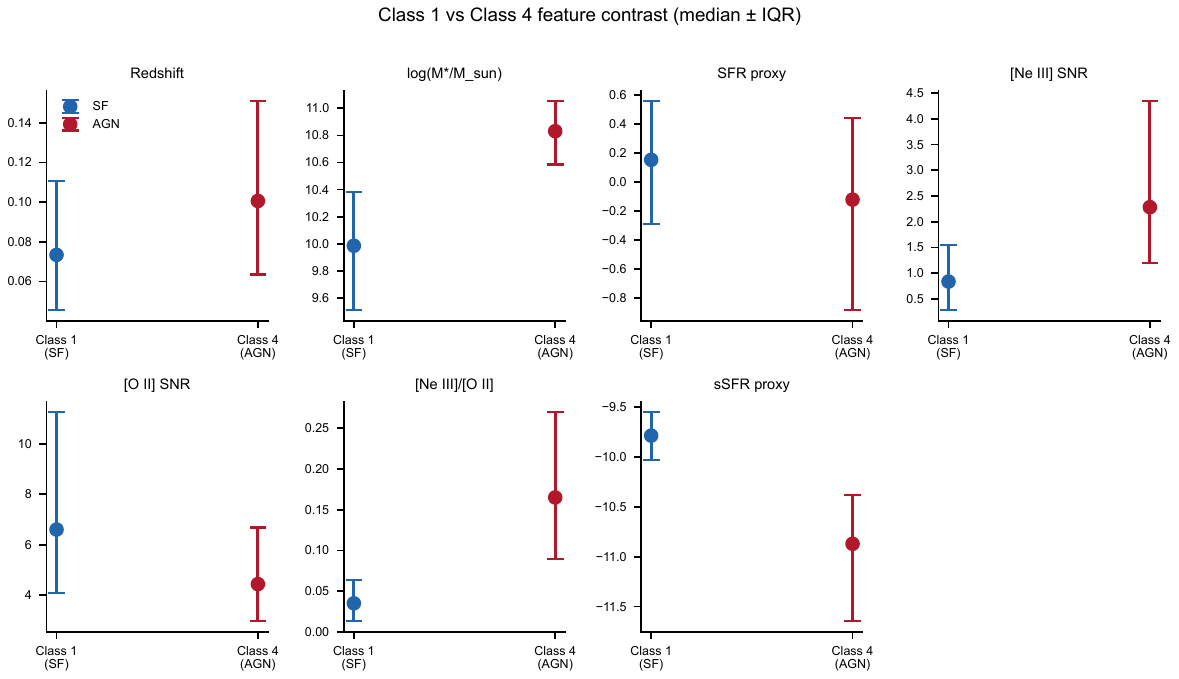}
\caption{
Comparison of feature distributions between Class~1 (star-forming
galaxies) and Class~4 (AGNs). AGN hosts tend to have higher stellar
masses, lower specific star-formation rates, and larger
\neiiioii\ ratios than star-forming galaxies. Several
missingness indicators, especially those associated with oxygen-abundance
measurements, also differ strongly between the two classes.
}
\label{fig:feature_contrast}
\end{figure*}
The correlation matrix in Figure~\ref{fig:feature_correlation} explains
why these rankings should be interpreted qualitatively rather than as a
strict ordering of independent physical causes. Several measured and
derived quantities form correlated groups, so importance can be shared
among variables that encode related information. At the same time, the
full feature set contains both correlated physical quantities and less
correlated availability flags. This mixture helps explain why the complete
27-dimensional representation outperforms models restricted to raw
measurements or to a smaller subset of derived quantities.

Figure~\ref{fig:feature_contrast} then compares the Class~1 and Class~4
feature distributions directly. The Class~1--Class~4 comparison in
Figure~\ref{fig:feature_contrast} makes the physical component of the
classification explicit. Relative to star-forming galaxies, AGN hosts are
shifted toward higher stellar mass, lower specific star-formation rate,
and larger Ne3O2 values. These trends match the established picture that
optically selected AGNs preferentially inhabit more massive, less strongly
star-forming hosts, while harder ionizing spectra enhance high-ionization
line ratios such as \neiiioii\ \citep{LevesqueRichardson2014}.
Thus, a substantial part of the learned Class~1--Class~4 separation is
anchored in physically interpretable host and emission-line differences,
not only in catalog artifacts.

\begin{figure}[!htbp]
\centering
\includegraphics[width=0.98\columnwidth,height=0.38\textheight,keepaspectratio]{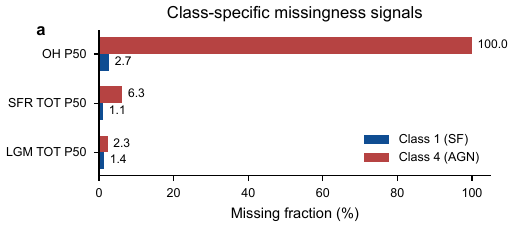}
\caption{
Comparison of feature-missingness rates between Class~1 (star-forming
galaxies) and Class~4 (AGNs).
}
\label{fig:missingness_contrast}
\end{figure}

Figure~\ref{fig:missingness_contrast} isolates the complementary catalog
signal carried by missingness indicators.
Figure~\ref{fig:missingness_contrast} shows the complementary catalog
signal. The strongest missingness contrast occurs for \texttt{OH\_P50},
the oxygen-abundance estimate in the MPA-JHU catalog. This contrast should
be interpreted as a catalog-dependent signal rather than as a direct
physical diagnostic. The MPA-JHU oxygen abundances are derived from
strong-line gas-phase metallicity estimates following
\citet{Tremonti2004}, so their availability depends on the relevant
emission-line measurements and on the catalog criteria used to assign
metallicities. The \texttt{OH\_P50\_missing} flag can therefore help the
classifier because it is coupled to line quality, BPT class, and catalog
construction, but it should not be read as independent physical evidence
for nuclear activity.

\begin{figure}[!htbp]
\centering
\includegraphics[width=0.98\columnwidth,height=0.38\textheight,keepaspectratio]{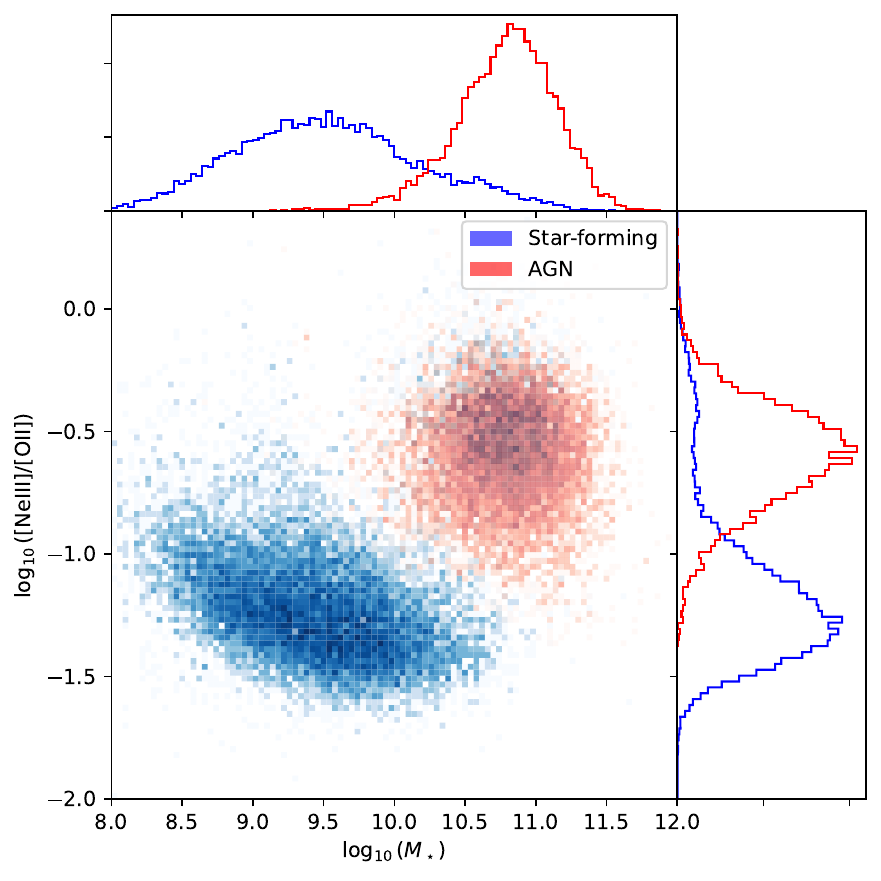}
\caption{
Distribution of galaxies in the \logneiiioii\ versus stellar mass plane.
}
\label{fig:ne3o2_mass}
\vspace{-0.8\baselineskip}
\end{figure}

The Ne3O2 diagnostic planes provide a final physical check on these
interpretation results. In Figure~\ref{fig:ne3o2_mass}, AGN hosts occupy
the high-mass, high-\logneiiioii\ region more frequently than
star-forming galaxies, while the star-forming population is concentrated
at lower Ne3O2 values. This pattern is consistent with the feature
rankings and the direct Class~1--Class~4 contrasts: Ne3O2 is informative
because it traces ionization conditions that differ systematically between
AGN hosts and star-forming systems.

\begin{figure}[!htbp]
\centering
\includegraphics[width=0.98\columnwidth,height=0.38\textheight,keepaspectratio]{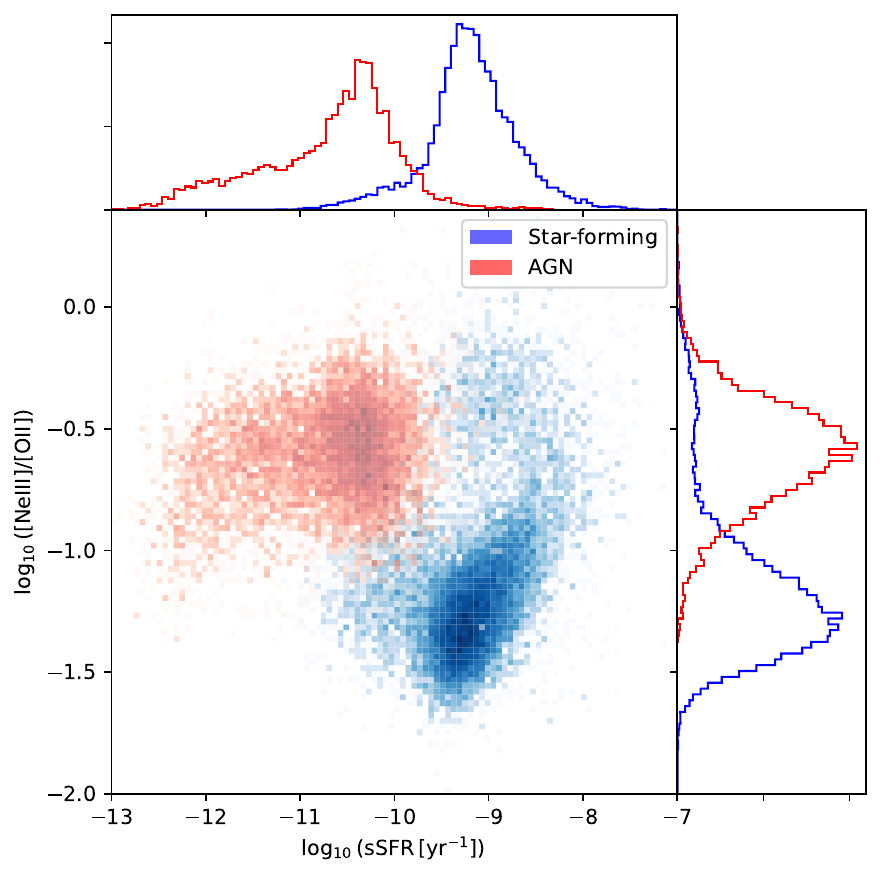}
\caption{
Distribution of galaxies in the \logneiiioii\ versus specific
star-formation rate plane.
}
\label{fig:ne3o2_ssfr}
\end{figure}

The \logneiiioii--sSFR plane in Figure~\ref{fig:ne3o2_ssfr} gives the
same conclusion from a complementary host-galaxy perspective. AGN hosts
are preferentially found at lower sSFR and higher \logneiiioii, whereas
star-forming galaxies occupy the higher-sSFR, lower-Ne3O2 region. Taken
together, Figures~\ref{fig:feature_importance}--\ref{fig:ne3o2_ssfr}
show that the models combine two kinds of information: physically
interpretable galaxy and line-ratio structure, and informative
missingness introduced by the construction of the spectroscopic catalog.
This combination is useful for ranking AGN candidates, but it also argues
against treating the classifier as a survey-independent replacement for
the BPT diagram without further validation.

\subsection{Implications for AGN Candidate Prioritization}

The preceding results indicate that the models are best interpreted as
AGN-ranking systems rather than hard replacements for physically
motivated BPT diagnostics. The Class~1 versus Class~4 ROC AUC of 0.9998
shows that the predicted probabilities contain substantial information
about AGN likelihood. However, the moderate hard-classification precision
for Class~4 shows that fixed decision thresholds can still introduce
contamination from adjacent BPT classes.

This distinction is important for survey applications. ROC-based metrics
alone can be misleading in highly imbalanced samples, so precision--recall
and top-$k$ retrieval analyses provide a more operationally relevant
assessment. When galaxies are ranked by predicted Class~4 probability,
the top 1\% of candidates reaches a precision of approximately 0.80 while
recovering about 44\% of all AGNs; expanding the list to the top 10\%
recovers nearly 97\% of the AGN population. Thus, compact high-purity
samples can support efficient follow-up, while broader lists can be used
when completeness is the priority.

We therefore argue that these models should be deployed as tools that
assign continuous AGN-likelihood scores for catalog construction and
target selection. This use is complementary to, rather than a substitute
for, traditional BPT diagnostics and subsequent physical interpretation,
particularly when standard BPT lines have low S/N or when rare AGN
populations must be identified efficiently in large spectroscopic survey
catalogs.

The results presented here should be interpreted in light of several limitations. First, all models were trained and evaluated on a single survey dataset. Their generalizability to other surveys with different selection functions,wavelength coverage, and noise characteristics remains to be established. Second, the class labels are themselves derived from the BPT framework that the models are designed to emulate. Consequently, any systematic biases or uncertainties inherent in the BPT classification scheme are propagated into the training targets. Third, although the adopted feature set extends beyond the traditional BPT inputs, it remains restricted to optical spectroscopic measurements. Incorporating photometric information or multiwavelength observations may further improve classification performance, particularly for the minority AGN-related classes. Finally, while we demonstrate that missingness patterns contain substantial predictive information, the underlying causal mechanisms remain unclear. Future work will be required to disentangle physically driven missingness, arising from the intrinsic properties of galaxies, from observational missingness introduced by survey design, measurement limitations, and catalog-construction procedures.

\section{Conclusions}
\label{sec:conclusions}

We have presented a survey-scale evaluation of machine-learning methods
for six-class BPT galaxy classification using 1,472,581 galaxies and a
27-dimensional feature representation. The analysis was designed to test
whether deep tabular architectures can improve upon strong classical and
gradient-boosted baselines, and whether missing-data patterns in
spectroscopic catalogs carry useful predictive information. The results
show that deep models are competitive, but their advantage over boosted
trees is small for the present low-dimensional tabular dataset. More
importantly, the predicted probabilities provide a useful ranking of AGN
candidates, suggesting that the practical value of these models lies in
prioritizing follow-up targets rather than producing uncontaminated hard
class labels. The main results are summarized as follows.

\begin{enumerate}

\item CNN--Transformer achieves the highest overall performance,
reaching a macro-F1 score of 0.6925, although the improvement over
LightGBM is modest. Deep tabular models and gradient-boosted decision
trees therefore occupy a similar performance regime for survey-scale
six-class BPT classification.

\item Missingness indicators provide substantial predictive power,
with the OH\_P50 missingness feature emerging as one of the
most informative variables. This finding suggests that survey
data-availability patterns encode observational or physical selection
effects that should be modeled explicitly rather than discarded.

\item Machine-learning models achieve very strong separation between
star-forming galaxies and AGNs in probability space
(Class~1 versus Class~4 AUC = 0.9998), even though six-class hard
classification remains affected by confusion among physically adjacent
BPT classes.

\item Feature-interpretation analyses show that the \logneiiioii\ ratio,
when combined with stellar mass or specific star-formation rate, provides
a physically informative separation between star-forming galaxies and AGN
hosts.

\item The models are best interpreted as AGN candidate-ranking tools
for large spectroscopic surveys rather than contamination-free
replacements for traditional BPT diagnostics. Precision--recall and
top-$k$ analyses show that the top 0.1\% of ranked candidates reaches a
precision of approximately 0.984, while the top 1\% maintains a precision
near 0.80 and recovers about 44\% of AGNs. Expanding the candidate list
to the top 10\% recovers nearly 97\% of the AGN population.

\end{enumerate}

In our future work, We will test the ranking calibration on independent surveys,
incorporate photometric or multiwavelength information, and examine
whether informative missingness remains stable under different selection
functions and catalog-generation pipelines. Such tests are necessary
before these models can be used as transferable AGN-prioritization tools
across heterogeneous spectroscopic surveys.

\section{Acknowledgments}

This work was supported by the China Scholarship Council--German
Academic Exchange Service (CSC--DAAD) joint scholarship program. The
computations in this study made use of the Gravity Supercomputer at the
Department of Astronomy, Shanghai Jiao Tong University. This work also
made use of computing resources provided by the Max Planck Computing and
Data Facility (MPCDF) through the Max Planck Institute for Astrophysics
(MPA).

S.G. acknowledges the Sloan Digital Sky Survey (SDSS), a wide-field optical
imaging and spectroscopic survey that provides homogeneous photometric
and spectroscopic data for large samples of astronomical objects. S.G. also acknowledges the MPA-JHU value-added galaxy catalog, which
augments SDSS galaxy spectra with emission-line measurements, stellar
masses, star-formation rates, metallicity estimates, and BPT
classifications that are central to the analysis presented here.

\bibliographystyle{aa}
\bibliography{aa}

\end{document}